\documentclass[12pt,preprint]{aastex}
\begin{document}
\title{ABELL 2744 MAY BE A SUPERCLUSTER ALIGNED ALONG THE SIGHTLINE}
\author{Jounghun Lee}
\affil{Astronomy Program, Department of Physics and Astronomy, Seoul National University, 
Seoul 08826, Republic of Korea} 
\email{jounghun@astro.snu.ac.kr}
\begin{abstract}
To explain the unusual richness and compactness of the Abell 2744, we propose a hypothesis that it may be a rich supercluster aligned along 
the sightline, and present a supporting evidence obtained numerically from the MultiDark Planck 2 simulations with a linear box size of 
$1\,h^{-1}$Gpc. 
Applying the friends-of-friends (FoF) algorithm with a linkage length of $0.33$ to a sample of the cluster-size halos from the simulations, 
we identify the superclusters and investigate how many superclusters have filamentary branches that would appear to be similar to the 
Abell 2744 if the filamentary axis is aligned with the sightline. Generating randomly a unit vector as a sightline at the position of the core 
member of each supercluster and projecting the positions of the members onto the plane perpendicular to the direction of the sightline, we 
measure two dimensional distances ($R_{2d}$) of the member halos from the core for each supercluster. Defining a Abell 2744-like 
spuercluster as the one having a filamentary branch composed of eight or more members with $R_{2d}\le 1\,$Mpc and masses comparable 
to those of the observed Abell 2744 substructures, we find one Abell 2744-like supercluster at $z=0.3$ and two at $z=0$. Repeating the same 
analysis but with the data from the Big MultiDark Planck simulations performed on a larger box of linear size of $2.5\,h^{-1}$Mpc,  we find that 
the number of the Abell 2744-like superclusters at $z=0$ increases up to eighteen, among which three are found more massive than 
$5\times 10^{15}\,M_{\odot}$.
\end{abstract}

\section{INTRODUCTION}\label{sec:intro}

The Abell 2744 cluster is such a rare event with virial mass approximately of $3\times 10^{15}\,M_{\odot}$ observed at $z\approx 0.308$, 
having a massive core substructure, from which seven distinct substructures are located $1\,$Mpc or less away in the plane of sky 
\citep[][and references therein]{AL_inferred}.  The combined analyses based on the strong and weak gravitational lensing effects revealed that 
the eight substructures (including the core) have typical cluster masses equal to or larger than $5\times 10^{13}\,M_{\odot}$ 
\citep{merten-etal11,medezinski-etal16,AL16}. 
Given that the fierce tidal effect in the central region of a rich cluster like the Abell 2744 would strip off its substructures, the presence 
of a throng of massive substructures at distances $\le 1\,$Mpc from its core in the Abell 2744 has been regarded quite anomalous  
\citep{AL16,AL_inferred}.

To quantify how improbable it is to find a cluster like the Abell 2744 in the standard $\Lambda$-dominated cold dark matter ($\Lambda$CDM)  
cosmology, \citet{AL_inferred} analyzed the catalog of dark matter (DM) halos from the Millennium XXL simulations with a linear box size of 
$3\,h^{-1}$Gpc \citep{mill_xxl} and looked for a Abell 2744-like cluster that is defined as a DM halo consisting of eight or more substructures 
with masses and distances from the core comparable to those of the Abell 2744. 
First of all, they attempted to make a parallel comparison between the observational and the numerical estimates of the masses of the 
substructures of the Abell 2744. What was measured from the gravitational lensing analyses is the partial mass enclosed 
in a cylinder with a radius of $150\,$kpc and an axis aligned with a given sightline.  
Whereas, what is available from the Millennium XXL simulation is the SUBFIND mass computed as the sum of DM particle masses belonging 
to each substructure identified by the SUBFIND algorithm \citep{subfind}.  
Assuming that the DM density of each substructure follows the Navarro-Frenk-White (NFW) profile \citep{nfw}, \citet{AL_inferred} inferred its 
SUBFIND mass with the constraint that its partial mass enclosed within a cylinder with a radius of $150\,$kpc matches the observational 
estimate, and looked for the Abell 2744-like clusters. 

Their searches for the Abell 2744-like clusters, however, turned out to be unfruitful with the data from the Millennium XXL simulations at 
$z=0.3$.  Noting that no systematics is likely responsible for this failure of finding a Abell 2744-like cluster and pointing out that taking into 
account baryonic proccesses would not alleviate the tension, \citet{AL_inferred} suggested that the structural properties of the Abell 2744  
be hardly explainable within the standard picture of structure formation based on the $\Lambda$CDM cosmology unless there exists some 
unknown mechanism that is capable of effectively debilitating the tidal stripping forces inside the Abell 2744. 

In this Paper, to explain more naturally the unusual compactness and richness of the Abell 2744 without appealing to an exotic physical process, we propose a new hypothesis that the Abell 2744 is not a gravitational bound cluster but a filamentary section of a marginally bound rich 
supercluster aligned along the sightline.  Consider a rich supercluster that has a thin filamentary branch composed of eight or more member 
clusters. If the direction of a sightline happens to coincide with the longest filamentary axis, then the filamentary branch of the supercluster 
might appear as a compact rich cluster like the Abell 2744.  
To numerically back up our hypothesis and to examine how probable it is to find such a supercluster with a filamentary branch composed of 
eight or more rich clusters,  we will make use of the data drawn from the MultiDark Planck 2 MDPL2 simulation \citep{MDPL} that are available 
from the CosmoSim database \footnote{https://www.cosmosim.org} and search for the Abell 2744-like {\it superclusters}.

\section{ABELL 2744-LIKE SUPERCLUSTERS}\label{sec:mdpl}

The MultiDark Planck 2 (MDPL2) simulation employed the L-Gadget2 code \citep{gadget} to track the trajectories of $3840^{3}$ dark matter 
particles from $z=120$ to $z=0$ in a periodic box of linear size $L_{box}=1\,h^{-1}$Gpc \citep{MDPL}.  It has force and mass resolutions of 
$5\,h^{-1}$kpc and $1.51\times 10^{9}\,h^{-1}M_{\odot}$, respectively,  and adopts the Planck $\Lambda$CDM cosmology \citep{planck14} 
to describe its initial conditions as 
$\Omega_{m}=0.307115,\ \Omega_{\Lambda}=0.692885,\Omega_{b}=0.048206,\ \sigma_{8}=0.8228,\ n_{s}=0.96,\ h=0.6777$.
The CosmoSim database released the catalogs of the DM halos identified at various snapshots via the application of the Rockstar halo finder 
\citep{rockstar} to the MDPL2 simulations, from which information on the key properties of each DM halo such as its virial mass, 
position and velocity can be extracted \citep{MD_release}. 

In fact, the catalog provides two different virial masses for each halo: one is computed as the sum of the masses of all DM particles located 
within the virial radius from the halo center, while the other is computed in the same way but by excluding the unbound DM particles within the 
virial radius from the sum. To make a parallel comparison with the observational estimates of the masses of the Abell 2744 cluster and its 
substructures that were made by the gravitational lensing analyses without excluding the unbound particles, we use the former definition for 
the viral masses of DM halos.  Throughout this Paper, the mass of a Rockstar halo denoted by $M_{h}$ refers to the total mass summed 
over all DM particles including the unbound ones within the virial radius. Moreover, to be consistent with the observational data, we express the 
masses and distances of the Rockstar halos in units of $M_{\odot}$ and $\,$Mpc, respectively, although MDPL2 catalog uses the  
units of $h^{-1}\,M_{\odot}$ and $h^{-1}\,$Mpc.
 
Applying the lower-mass cut of $M_{h,cut}=5\times 10^{13}\,M_{\odot}$ to the Rockstar halo catalog at $z=0.3$, the redshift at which 
the Abell 2744 is observed from the southern hemisphere \citep{abell89}, we construct a sample of the cluster-size halos, to which the 
friends-of-friends algorithm (FoF) with a linking length of $l_{c}\equiv \bar{l}/3$ is applied, where $\bar{l}$ is the mean separation distance 
of the cluster-size halos.  
Using the FoF groups of the cluster-size halos as the supercluster proxies, we compute the total mass of each supercluster, $M_{sc}$, as 
$M_{sc}=\sum_{i=1}^{N_{h}}M_{h,i}$ where $M_{h,i}$ is the mass of the $i$th member and $N_{h}$ is the number of the member halos.  
Imposing the conditions of $M_{sc}\ge M_{sc,cut}=3\times 10^{15}\,M_{\odot}$ and $N_{h}\ge N_{h,cut}=8$ on the superclusters, 
we select a total of $33$ candidates for the Abell 2744-like superclusters at $z=0.3$ . 
It is worth mentioning here that the cut-off values, $M_{h,cut}$ , $M_{sc,cut}$ and $N_{h,c}$, are deliberately chosen to match the properties 
of the Abell 2744 whose $N_{h,cut}$ substructures are more massive than $M_{h,cut}$, having a total mass of $\sim M_{sc,cut}$
\citep[][and references therein]{AL_inferred}

Designating the most massive member halo as its core for each candidate supercluster, we calculate the separation vectors of the 
member halos relative to the core as ${\bf r}={\bf x}-{\bf x}_{c}$ where ${\bf x}$ and ${\bf x}_{c}$ are the position vectors of the member 
and the core halos, respectively.  At the location of the core of each candidate, we project the separation vectors ${\bf r}$ onto a two 
dimensional plane perpendicular to a randomly generated unit vector and count the member halos whose projected separation distances, 
say $R_{2d}$, from the core are  equal to or less than $1\,$Mpc.
If a candidate supercluster is found to have eight or more member halos with $R_{2d}\le 1\,$Mpc in the projected plane along a certain 
sightline, then it is selected as a Abell 2744-like supercluster.  The algorithm via which we search for the Abell 2744-like superclusters 
among the candidates is presented in the following.
\begin{enumerate}
\item
Generate two independent random numbers, say $\theta$ and $\phi$, in the range of $[0, 1]$ and multiply them by $\pi$ and $2\pi$, 
respectively.  Construct a unit vector ${\bf u}$ from $\theta$ and $\phi$ such as 
${\bf u} = \left(\sin\theta\cos\phi,\sin\theta\sin\phi,\cos\theta\right)$. 
\item
For a given candidate supercluster, project the separation vectors ${\bf r}$ of its member halos onto a plane perpendicular to the unit vector 
${\bf u}$, and then calculate the projected separation distances as $R_{2d}\equiv \vert{\bf r}-\left({\bf r}\cdot{\bf u}\right){\bf u}\vert$. 
\item
Count the member halos which satisify the condition of $R_{2d}\le 1\,$Mpc. If the counted number, $N_{h}(R_{2d}\le 1\,{\rm Mpc})$, is equal to 
or larger than eight, then the candidate is selected as a Abell 2744-like supercluster.
\item
If $N_{h}(R_{2d}\le 1\,{\rm Mpc})$ is less than eight, then return to step 1 and redo the calculations  with a newly generated 
random direction ${\bf u}$. 
\item
If a candidate never satisfies the condition of $N_{h}(R_{2d}\le 1\,{\rm Mpc})\ge 8$ while the steps 1-4 are repeated 1000 times, 
then it is ruled out.
\end{enumerate}

\begin{figure}
\begin{center}
\plotone{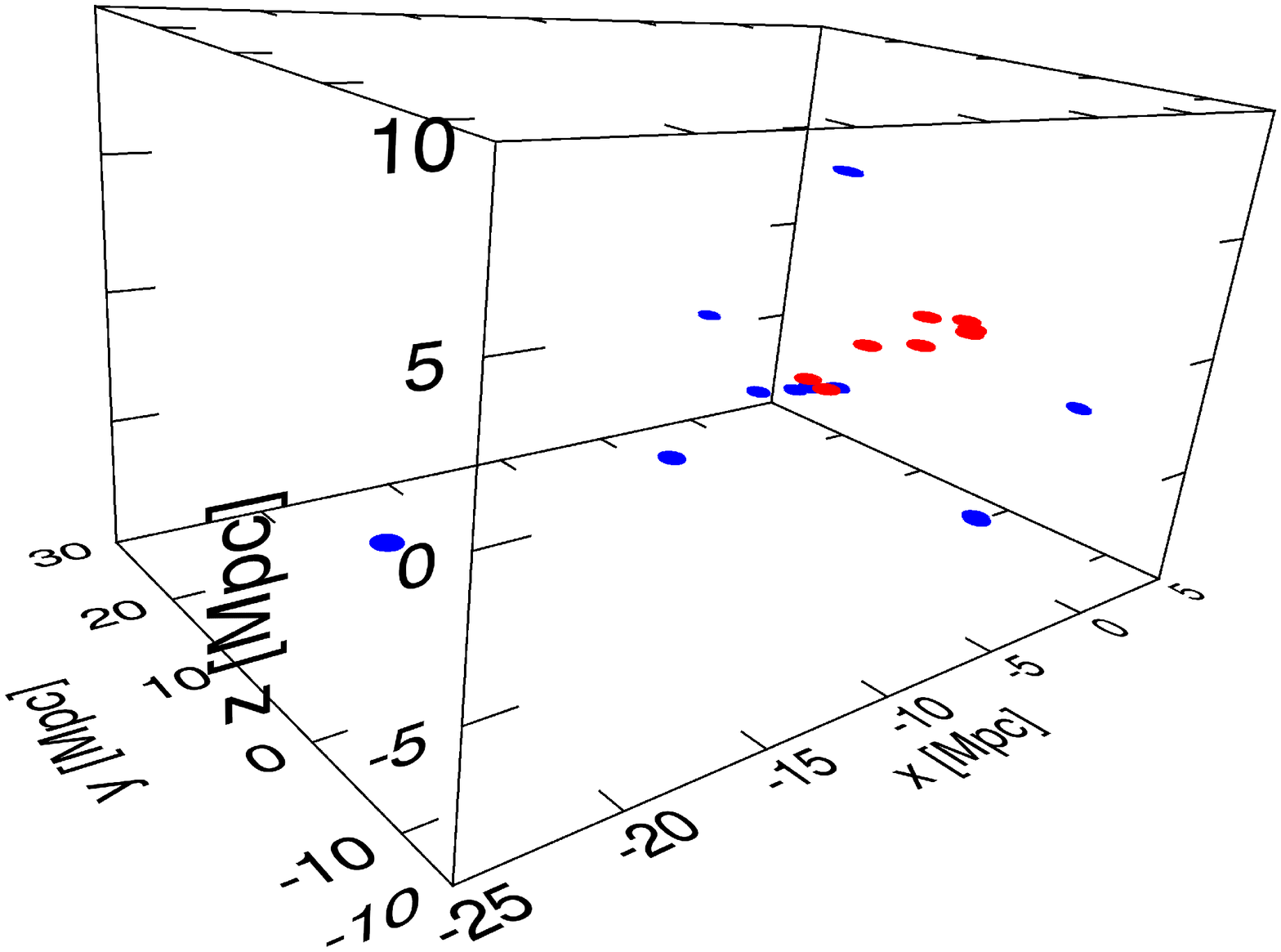}
\caption{Three dimensional spatial distribution of all member halos belonging to the Abell 2744-like supercluster 
detected at $z=0.3$ from the MDPL2 simulation.}
\label{fig:viewall3d}
\end{center}
\end{figure}

\begin{figure}
\begin{center}
\plotone{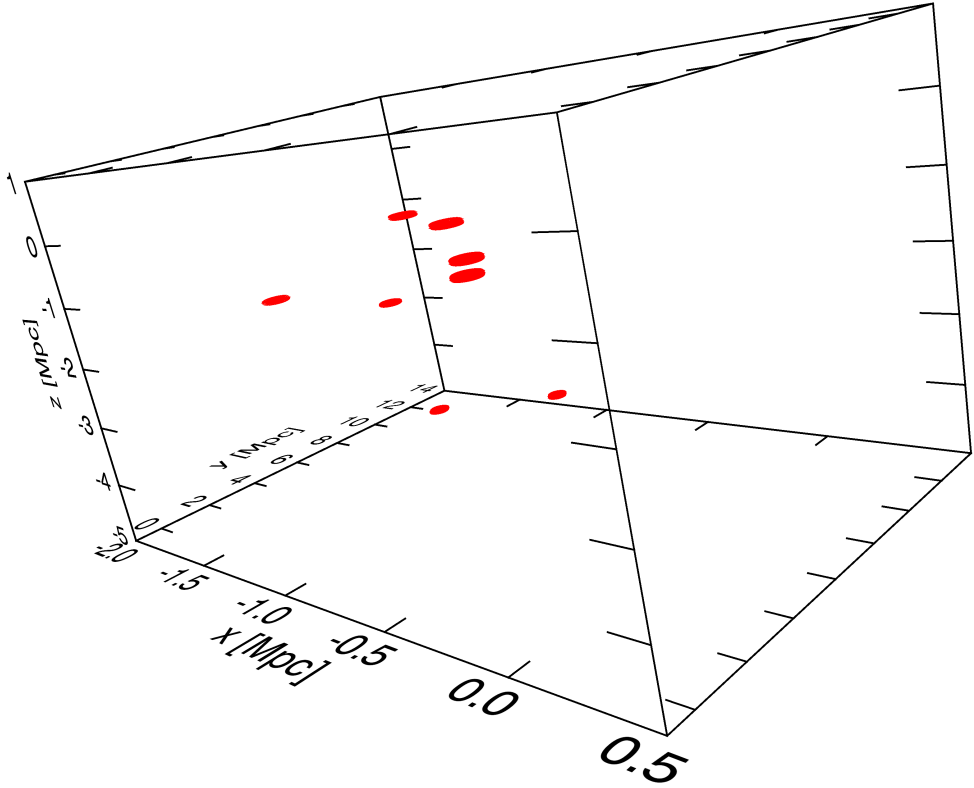}
\caption{Same as Figure \ref{fig:viewall3d}  but zoomed in on the the eight member halos belonging to the filamentary 
branch from the direction of the filamentary axis.}
\label{fig:view3d}
\end{center}
\end{figure}

By following the above procedures with the $33$ candidates selected at $z=0.3$, we detect only one as a Abell-2744 like supercluster, which 
is found to have a total mass of $M_{sc}=5.65\times 10^{15}\,M_{\odot}$  and $19$ member halos.  Its filamentary branch that consists of  
$8$ members with $R_{2d}\le 1\,$Mpc in the plane perpendicular to the direction of its filamentary axis is found to have a total mass of 
$3.59\times 10^{15}\,M_{\odot}$. 
Figure \ref{fig:viewall3d} shows three dimensional spatial locations of the member clusters relative to the core of the detected 
Abell 2744-like supercluster at $z=0.3$ from the MDPL2 simulation. The red dots correspond to the eight members of the filamentary branch 
while the blue dots are the other members. Figure \ref{fig:view3d} zooms in on the spatial distributions of the eight members viewed from 
the direction of its longest filamentary axis.

\begin{figure}
\begin{center}
\plotone{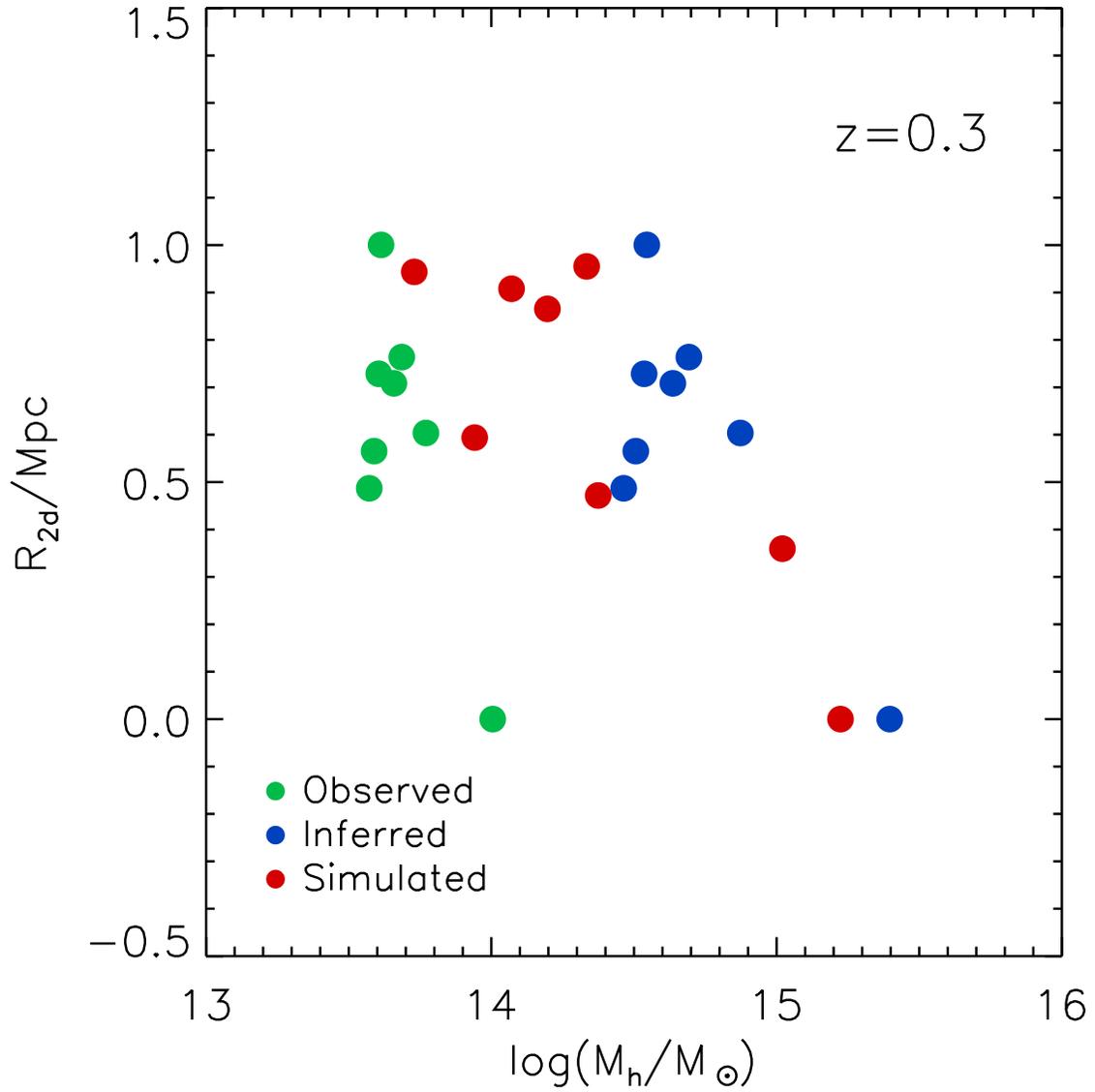}
\caption{Configurations of the observed, inferred and simulated members of the Abell 2744 in the 
plane spanned by the mass and the projected distance (green, blue and red dots, respectively). }
\label{fig:mdis_z0.3}
\end{center}
\end{figure}

Figure \ref{fig:mdis_z0.3} depicts the configurations of the eight members belonging to the filamentary branch of the selected 
Abell 2744-like supercluster at $z=0.3$ as red dots in the plane spanned by $M_{h}$ and $R_{2d}$, and compare them with the observed 
and inferred configulrations of the real Abell 2744 substructures (green and blue dots, respectively). As mentioned in Section \ref{sec:intro}, 
the lensing observations estimated the partial masses of the Abell 2744 substructures within a cylindrical radius of $150\,$kpc 
\citep{merten-etal11,medezinski-etal16,AL16}. The green dots in Figure \ref{fig:mdis_z0.3} correspond to these partial masses of the 
substructures versus their projected distances in the plane of the sky.
The blue dots in Figure \ref{fig:mdis_z0.3} correspond to the {\it inferred} virial masses versus the projected distances. 
\citet{AL_inferred} inferred the SUBFIND masses of the Abell 2744 substructures from their observed partial masses with the help of the 
NFW profile, where the SUBFIND masses are not the virial masses but $1.34$ times higher (J. Schwinn 2017, private communication).
To make a parallel comparison with our results that utilize the virial masses of the member halos of the Abell 2744-like superclusters, 
we divide the SUBFIND masses of the Abell 2744 substructures by $1.34$ to obtain their virial masses. 
From here on, the observed and the inferred masses of the Abell 2744 substructures refer to the partial masse estimated 
by the lensing observations and the virial masses inferred by \citet{AL_inferred} (SUBFIND masses divided by $1.34$), respectively. 
Meanwhile the virial masses of the eight members belonging to the filamentary branch of the Abell 2744-like supercluster detected in the 
current work will be called the simulated ones.

Figure \ref{fig:mdis_z0.3} displays a notable difference between the inferred and the simulated masses. The distribution of the latter is 
spreaded over a wide range of $13.6\le \log (M_{h}/M_{\odot})\le 15.3$ while that of the former is more inclined toward the higher mass section 
of $\log (M_{h}/M_{\odot})\ge 14.3$. This difference between the two results, however, is likely due to the over-prediction of the virial masses 
of the Abell 2744 substructures, as admitted by \citet{AL_inferred}.  In fact, it can be easily verified that \citet{AL_inferred} over-estimated 
the SUBFIND masses of the Abell 2744 substructures by examining the total mass of the Abell 2744, which has been known to be around 
$3\times 10^{15}\,M_{\odot}$ \citep{AL_inferred}. The sum of the virial masses (SUBFIND masses divided by 1.34) inferred by 
\citet{AL_inferred} is far above this total mass, reaching up to $6\times 10^{15}\,M_{\odot}$, which obviously indicates that their values 
are over-estimated (J. Schwinn 2017, private communication).

It is worth emphasizing here that what we have detected as the Abell 2744 supercluster at $z=0.3$ from the MDPL2 simulation is not a structure 
having the same structural properties as the Abell 2744 but a structure which has a filamentary branch that could look similar to the Abell 2744 
only if the sightline happens to coincide with the direction of the filamentary axis. If it is viewed from a different direction, then it would not 
appear similar to the Abell 2744.  While \citet{AL_inferred} tried only in vain to find a structure as rich and compact as the Abell 2744 from the 
Millennium XXL simulations run on a box of linear size of $3\,h^{-1}\,$Gpc,  we have find one supercluster which could appear similar to 
the Abell 2744 from the MDPL2 simulations run on a box of three times smaller linear size. This result is found to be robust against 
slight variations of $l_{link}$ and $M_{sc}$.

\begin{figure}
\begin{center}
\plotone{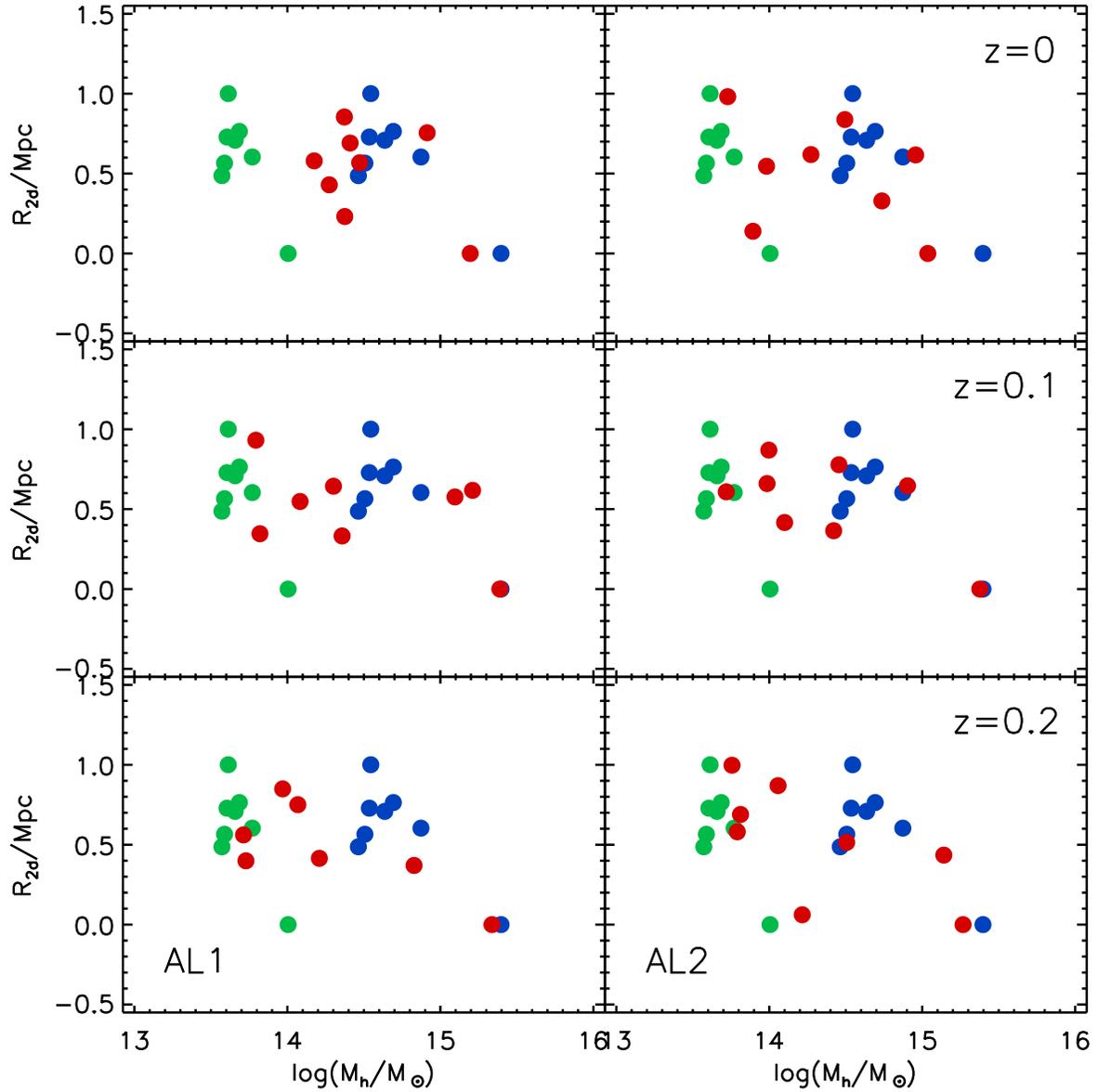}
\caption{Same as Figure \ref{fig:mdis_z0.3} but at $z=0.2,\ 0.1\ 0$ (in the bottom, middle and top panels, respectively). 
At each redshift, two Abell 2744-like superclusters are identified and marked as AL1 and AL2.}
\label{fig:mdis}
\end{center}
\end{figure}

To see how the number of the Abell 2744-like superclusters, say $N_{AL}$, changes with the decrement of $z$, we apply the above 
algorithm to the Rockstar halo catalogs at three different redshifts, $z=0,\ 0.1,\ 0.2$ and search for the Abell 2744-like superclusters. 
At each redshift are detected two Abell 2744-like superclusters (say, AL1 and AL2), whose $M_{h}$-$R_{2d}$ configurations are plotted
as red dots in Figure \ref{fig:mdis} and compared with the inferred and the observed values (the blue and the green dots, respectively). 
As can be seen, at $z=0,1$ and $0.2$, the simulated and inferred masses of the core halos (with $R_{2d}=0$) agree quite well with each other, 
while at the present epoch ($z=0$), the simulated masses of the cores of the AL1 and AL2 are lower than the inferred values.  

To see how $N_{AL}$ increases with the increment of the simulation box size, we use the Rockstar halo catalog at $z=0$ from the 
BigMultiDark Planck (BigMDPL) that is the same as the MDPL2 but performed in a larger box of linear size $L_{box}=2.5\,h^{-1}$Gpc. 
Since only the $z=0$ snapshot data is available from the BigMDPL simulation at the CosmoSim database, we apply the above algorithm 
to the Rockstar halo catalog at $z=0$ from the BigMDPL and detect a total of eighteen Abell 2744-like superclusters.  
Table \ref{tab:nsc} lists the number of the candidate superclusters ($N_{sc}$) and the number of the detected Abell 2744-like superclusters 
for the four different cases of $L_{box}$ and $z$. 

\begin{figure}
\begin{center}
\plotone{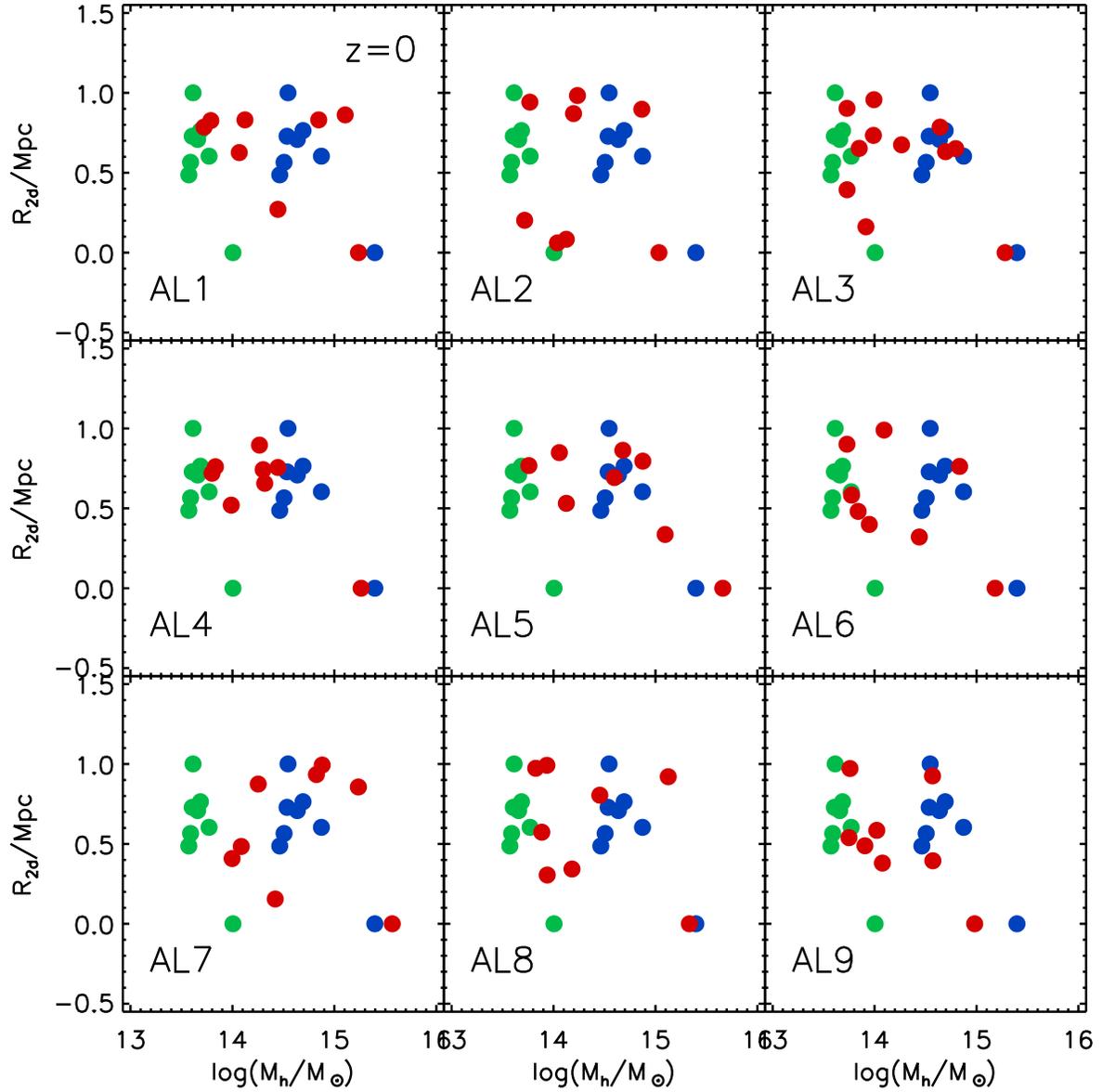}
\caption{Same as Figure \ref{fig:mdis_z0.3} but for the cases of nine Abell 2744-like superclusters from the 
BigMDPL simulation at $z=0$.} 
\label{fig:mdis_big1}
\end{center}
\end{figure}

\begin{figure}
\begin{center}
\plotone{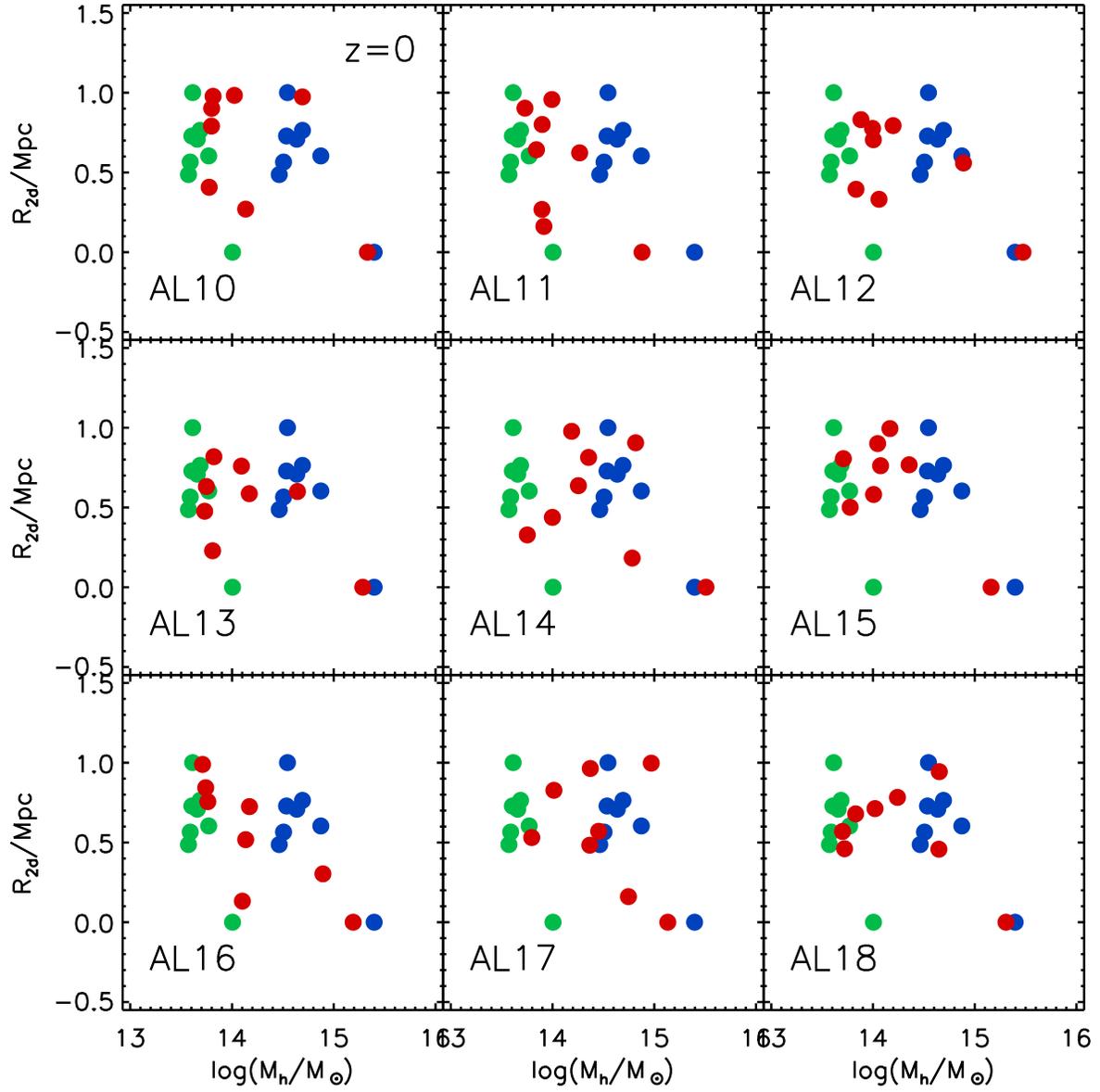}
\caption{Same as Figure \ref{fig:mdis_big2} but the other nine Abell 2744-like superclusters.} 
\label{fig:mdis_big2}
\end{center}
\end{figure}

Each of panel in Figures \ref{fig:mdis_big1} and \ref{fig:mdis_big2} shows the $M_{h}$-$R_{2d}$ configurations of the member halos 
belonging to the filamentary branch of each Abell 2744-like supercluster detected at $z=0$ from the BigMDPL simulation (red dots), and 
compare them with the inferred and the observed configurations (the blue and the green dots, respectively). 
As can be seen, three of the eighteen Abell 2744-like superclusters have the cores more massive than $3\times 10^{15}\,M_{\odot}$ 
(AL5, AL7, AL14).  For each of the eighteen Abell 2744-like superclusters found in the BigMDPL simulation , the number of all of their 
member halos ($N_{sc}$), the number and mass of those member halos belonging to the filamentary branch ($N_{fc}$ and $M_{fc}$, 
respectively), and the mass of the core halo ($M_{core}$) are listed in Table \ref{tab:msub}. 

\section{SUMMARY AND DISCUSSION}\label{sec:con}

Hypothesizing that the Abell 2744 may not be a gravitationally bound cluster but a filamentary section of a marginally bound supercluster 
aligned along the sightline, we have investigated how probable it is to find a Abell 2744-like supercluster in a Planck Universe by analyzing the 
Rockstar halo catalogs at $z=0.3$ from the MDPL2 simulations with a linear box size of $1\,h^{-1}$Gpc.  
Using the FoF groups of the cluster-size halos with masses $M_{h}\ge 5\times 10^{13}\,M_{\odot}$ as the supercluster proxies,  
we have first found $33$ candidate superclusters which have eight or more member halos  with total masses of 
$M_{sc}\ge 3\times 10^{15}\,M_{\odot}$. Among the 33 candidates one supercluster is found to have a narrow filamentary branch composed of 
eight members with projected distances equal to or less than $1\,$Mpc from the most massive core in the plane perpendicular to a randomly 
selected sightline direction. 

Following the same procedure but with the data at $z=0$ from the BigMDPL simulation run on a box of linear size $2.5\,$Gpc 
has led us to detect eighteen Abell 2744-like superclusters. Recalling that \citet{AL_inferred} found no Abell 2744-like cluster from 
the Millennium XXL simulation run on a larger box \citep{AL_inferred}, we conclude that it is still rare but not impossible to detect a Abell 
2744-like structure in the Planck universe under our hypothesis. In other words, our hypothesis makes it less difficult to explain the observed 
structural properties of the Abell 2744 in the standard theory of structure formation based on the $\Lambda$CDM cosmology. 

A couple of follow-up works will be necessary to test further our hypothesis.  First, to make a fairer comparison between the numerical 
prediction and the observational data, the virial masses of the members of the Abell 2744 as well as its total mass should be estimated 
with higher accuracy and precision from real data since the values provided by the work of \citet{AL_inferred} turned out to be over-estimated.  
Second, our algorithm for the detection of the Abell 2744-like superclusters should be tested by applying it to a light-cone simulation 
to determine practically the probabilty of finding a structure similar to the Abell 2744. 
In the current analysis that has used one fixed snapshot data from the MDPL2 simulation, the direction of a line-of-sight had to be randomly 
generated randomly at the position of each supercluster core.  Thus,  what our detection of a Abell 2744-like supercluster really implies 
is that a filamentary branch of this supercluster has a {\it potential capacity} to look similar to the Abell 2744 and that this potential capacity 
is contingent upon the direction of a sightline. With a light-cone simulation data for which the line-of-sight direction are given for each 
supercluster, it will be possible to evaluate not the potential but the true probability of finding a Abell 2744-like structure in the Universe. 

We also speculate that our hypothesis might be able to explain the hot X-ray emissions of the Abell 2744 \citep{merten-etal11}. If the 
Abell 2744 is indeed a filamentary branch of a rich supercluster whose member clusters are in the middle of colliding onto the core, the 
narrow filamentary channel will play a role in speeding up the member clusters and thus increasing the intra-cluster temperature. It will 
be interesting to examine how the intra-cluster temperature increases in the filamentary environment like the Abell 2744 in the frame of 
our hypothesis. Our future work is in this direction. 

\acknowledgements

I acknowledge the support of the Basic Science Research Program through the NRF of Korea funded by the Ministry of Education 
(NO. 2016R1D1A1A09918491).  I was also partially supported by a research grant from the National Research Foundation (NRF) of Korea 
to the Center for Galaxy Evolution Research (NO. 2010-0027910). 

The CosmoSim database used in this paper is a service by the Leibniz-Institute for Astrophysics Potsdam (AIP).
The MultiDark database was developed in cooperation with the Spanish MultiDark Consolider Project CSD2009-00064.
The authors gratefully acknowledge the Gauss Centre for Supercomputing e.V. (www.gauss-centre.eu) and the Partnership for Advanced 
Supercomputing in Europe (PRACE, www.prace-ri.eu) for funding the MultiDark simulation project by providing computing time on the GCS 
Supercomputer SuperMUC at Leibniz Supercomputing Centre (LRZ, www.lrz.de).
The Bolshoi simulations have been performed within the Bolshoi project of the University of California High-Performance AstroComputing 
Center (UC-HiPACC) and were run at the NASA Ames Research Center.

\clearpage
\clearpage
\clearpage
\clearpage
\clearpage
\clearpage
\clearpage
\begin{deluxetable}{cccc}
\tablewidth{0pt}
\setlength{\tabcolsep}{5mm}
\tablecaption{Linear box size of the simulation, the redshift, the number of the supercluster candidates and the number of the 
Abell 2744-like superclusters}
\tablehead{$L_{box}$ & $z$ & $N_{sc}$ & $N_{AL}$ \\
($h^{-1}$Gpc) & & &}
\startdata
1  & $0$ & $90$ & $2$ \\
1 & $0.1$ & $58$ & $2$ \\
1 & $0.2$ & $40$ & $2$ \\
1 & $0.3$ & $33$ & $1$ \\
2.5 & $0$ & $1104$ & $18$ \\
\enddata
\label{tab:nsc}
\end{deluxetable}
\clearpage
\begin{deluxetable}{cccc}
\tablewidth{0pt}
\setlength{\tabcolsep}{5mm}
\tablecaption{Numbers of all member halos of the eighteen Abell 2744-like supercluster and the numbers of the member halos belonging 
to their filamentary branch , the masses of the filamentary branches and the masses of the cores}
\tablehead{$N_{sc}$ & $N_{fc}$ & $M_{fc}$ & $M_{core}$ \\
& & $(10^{15}\,h^{-1}M_{\odot}$) & $(10^{15}\,h^{-1}M_{\odot}$)}
\startdata
          34 &           8 &   4.35    &   1.73    \\
          28 &           8 &   2.50     &   1.08     \\
          27 &           8 &   3.87     &   1.91     \\
          26 &           8 &   2.93     &   1.83     \\
          24 &           8 &   7.76     &   4.58    \\
          21 &           8 &   2.87    &   1.52     \\
          20 &           8 &   7.50     &   3.70     \\
          19 &           8 &   4.24     &   2.15     \\
          18 &           8 &   2.12     &  0.96     \\
          18 &           8 &   3.13    &   2.14     \\
          17 &           8 &   1.41     &  0.76     \\
          16 &           8 &   4.37     &   2.98     \\
          15 &           8 &   2.88     &   1.93     \\
          12 &           8 &   5.20     &   3.22     \\
          11 &           8 &   2.26     &   1.45     \\
          11 &           8 &   2.91     &   1.55     \\
          11 &           8 &   3.77     &   1.36     \\
           9 &           8 &   3.37     &   2.03     \\
\enddata
\label{tab:msub}
\end{deluxetable}
\end{document}